\documentstyle[multicol,aps,prb,epsfig]{revtex}

\begin{document}

\title{Memory and chaos in an Ising spin glass}

\author{R. Mathieu, P. E. J\"onsson, and P. Nordblad}
\address{Department of Materials Science, Uppsala University, Box 534, SE-751 21 Uppsala, Sweden}

\author{H. Aruga Katori and A. Ito}
\address{RIKEN (The Institute of Physical and Chemical Research), Wako, Saitama, 351-0198, Japan}

\date{\today}

\maketitle

\begin{abstract}

The non-equilibrium dynamics of the model 3d-Ising spin glass - Fe$_{0.55}$Mn$_{0.45}$TiO$_3$ - has been investigated from the temperature and time dependence of the zero field cooled magnetization recorded under certain thermal protocols. The results manifest chaos, rejuvenation and memory features of the equilibrating spin configuration that are very similar to those observed in corresponding studies of the archetypal RKKY spin glass Ag(Mn). 
The sample is rapidly cooled in zero magnetic field, and the magnetization recorded on re-heating. When a stop at constant temperature $T_s$ is made during the cooling, the system evolves toward its equilibrium state at this temperature. The equilibrated state established during the stop becomes frozen in on further cooling and is retrieved on re-heating. The memory of the aging at $T_s$ is not affected by a second stop at a lower temperature $T'_s$. Reciprocally, the first equilibration at $T_s$ has no influence on the relaxation at $T'_s$, as expected within the droplet model for domain growth in a chaotic landscape.
\end{abstract}


\begin{multicols}{2}

\section{Introduction}

In the droplet model of spin glasses (SG)\cite{fisher}, the equilibrium dynamics is governed by low energy excitations from the ground state (droplets). If a weak magnetic field - weak enough not to affect the system - is applied, the equilibrium system is magnetized via polarization of 
droplets of ever increasing size. However,
the droplet model SG phase is chaotic\cite{bray} in the sense that the equilibrium configuration is unstable to any temperature change $\Delta T$ (or change in the distribution of the interaction) and thus has to rearrange after such a perturbation. For example, after a quench from the paramagnetic phase into the spin glass phase the system is trapped in a random non-equilibrium spin configuration which slowly equilibrates (ages). The aging occurs via droplet excitations which equilibrate the spin configuration on larger and larger length scales, causing the spin glass correlation length to grow. Also, the equilibrium configurations at two different temperatures $T$ 
and $T+\Delta T$ are only different on large length scales, outside an overlap region.\\
It is observed experimentally that if an equilibrium SG configuration is imprinted up to a  certain length scales (age) by keeping the spin glass a certain time at constant 
temperature, this length scale is conserved even though new configurations on shorter length scales are imprinted at other (lower) temperatures. Manifestations of memory and chaos in spin glasses have 
been observed in the relaxation and temperature dependence of the low frequency 
ac-susceptibility in thermal protocols containing halts at constant temperature\cite{acmemo,memo}, as well as in temperature cycling experiments\cite{cycl}. In addition, signatures of 
memory and chaos 
have been demonstrated in simple dc-magnetization vs. temperature measurements\cite{trmpap} on 
a standard RKKY spin glass, Ag(Mn). 
In the present report the  
model 3d Ising spin-glass, Fe$_{0.55}$Mn$_{0.45}$TiO$_3$, is investigated using 
dc-magnetization experiments. 
The temperature dependence, as well as the relaxation of the zero field cooled (ZFC) 
magnetization is recorded after specific cooling protocols. The results evidence memory and 
chaos phenomena in agreement with the predictions from the droplet model for the chaotic SG 
phase.   

\section{Experimental}

The ZFC magnetization of a single crystal of Fe$_{0.55}$Mn$_{0.45}$TiO$_3$ was recorded versus temperature and time in a non-commercial low-field SQUID magnetometer\cite{SQUID}. This spin 
glass is considered as a canonical short-range Ising spin-glass\cite{Ito}, with a SG transition temperature $T_g$ $\sim$ 22 K. In the experiments, the cooling of the sample was 
interrupted by stop(s) at constant temperature. In the case of single stop experiments (SSE), 
the sample was first rapidly cooled in zero magnetic field from a reference temperature $T_{ref}$ $>$ $T_g$ down to the stop temperature $T_s$ $<$ $T_g$, where it was kept for $t_s$=3000s 
(and other durations). 
The cooling was then resumed down to the lowest temperature where the magnetic field $h$=1 Oe 
was applied, and the magnetization recorded on re-heating. In the double stops experiments 
(DSE) case, an equally long  second temperature stop was made at $T'_s$ $<$ $T_s$. A reference 
curve was recorded after direct cooling of the sample to the lowest temperature.\\
The same cooling protocols were employed in the relaxation measurements. After re-heating to 
the measurement temperature $T_m$, and a wait time of $t_w$=10s, a magnetic field $\Delta h$=2 
Oe was applied, and the magnetization recorded vs. time elapsed after the field application. 
In the SSE protocol, a 3000s stop was made a $T_s$=$T_m$ - the 
experiment is thus comparable 
to a short negative temperature cycling - while in the DSE, two 3000s stops were made at $T_s$ 
and $T'_s$ $<$ $T_s$, and the relaxation measured at both $T_m$=$T_s$ and $T_m$=$T'_s$. 
Reference measurements (RE) were performed at $T_m$ by cooling the sample directly to the 
lowest temperature, and recording $M$ vs. time in $\Delta h$=2 Oe after re-heating the sample 
to $T_m$ and employing the wait time of $t_w$=10s. ``Conventional'' ZFC relaxation 
measurements, in which the sample was only rapidly cooled to $T_m$, with $t_w$=3000s, 
were performed for comparison and to ensure that the magnitude of the field was low 
enough to give a linear response\cite{mattsson} from the system. \\
The dc magnetic field was applied along the $c$-axis of the single-crystal. It is generated 
by a small superconductive solenoid coil working in persistent mode during measurements. 
The time constant of the superconducting magnet is of order 1 ms, but the whole procedure to 
change the field from $h$ to $h+\Delta h$ and re-establish persistent mode in the magnet takes 
$\sim$ 0.2 s which also determines the shortest observation time in the relaxation experiments.
Apart from the stops at constant temperature, the cooling rate was $\sim$ 4K/min, and the 
heating rate of the order of 1K/min.

\section{Results and discussion}

\begin{figure}
\centerline{\epsfig{figure=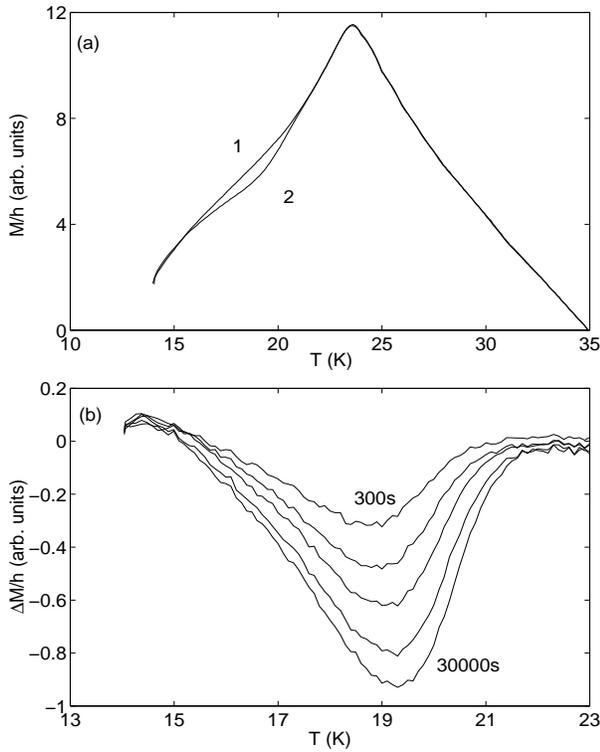,height=10cm,width=8cm}}
\caption{(a) ZFC magnetization vs. temperature. Curve {\bf 1} is a reference curve, while curve {\bf 2} show the effects of a 3000s stop at $T_s$=19K during the initial cooling. In (b) curve {\bf 2} is subtracted from the the reference; the results are shown for different waiting times, logarithmically spaced from 300 to 30000s.}
\label{fig1}
\end{figure}

\begin{figure}
\centerline{\epsfig{figure=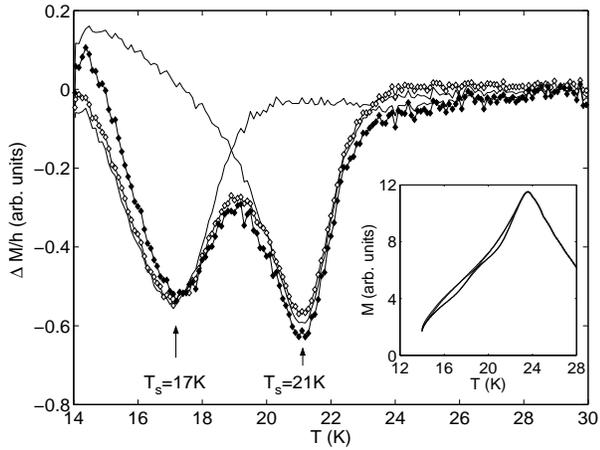,height=6cm,width=8cm}}
\caption{Difference plots corresponding to single-stops of 3000s at $T_s$=17K and $T_s$=21K (simple lines), and double-stops of 3000s both at $T_{s1}$=21K and $T_{s2}$=17K (open diamonds). The sum of the two single-stop curves is added for comparison (filled diamonds). The insert shows the original ZFC magnetization vs. temperature curves.}
\label{fig2}
\end{figure}

\begin{figure}
\centerline{\epsfig{figure=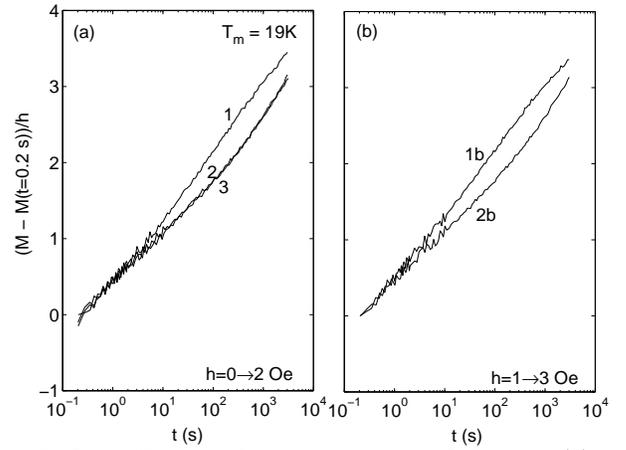,height=6cm,width=8cm}}
\caption{ZFC relaxation curve for $\Delta h$=2 Oe. In (a) the sample is first cooled in zero magnetic field to the lowest temperature and re-heated to the measurement temperature $T_m$=19K. After $t_w$=10s, $\Delta h$=2 Oe is applied and the magnetization recorded vs. time. Curve {\bf 1} is a reference curve measured without a stop during cooling, while curve {\bf 2} shows the effects of an additional stop of 3000s at $T_s$=19K during cooling. Curve {\bf 3} corresponds to a conventional ZFC relaxation experiment, with cooling down to the measurement temperature only, with $t_w$=3000s. (b) shows the same experiments, performed this time cooling the sample is a field of $h$=1 Oe, and recording the magnetization in $h$=3 Oe.}
\label{fig3}
\end{figure}

\begin{figure}
\centerline{\epsfig{figure=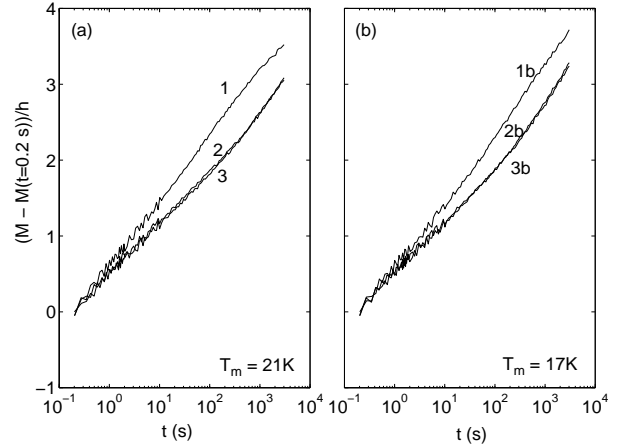,height=6cm,width=8cm}}
\caption{Comparison to the double-stop experiments in $M$ vs. $T$. Curve {\bf 1} and {\bf 2} (resp. {\bf 1b} and {\bf 2b}) correspond to the relaxation measurements described in Fig. 3(a), for (a) $T_m$=21K and (b) $T_m$=17K. In the case of {\bf 3} (resp.{\bf 3b}), a first 3000s stop in the initial cooling is made at $T_{s1}$=21K before the second one at $T_{s2}$=17K.} 
\label{fig4}
\end{figure}

Figure~\ref{fig1} illustrates the effects on the ZFC magnetization of a stop at constant 
temperature during the cooling. As seen in Fig.~\ref{fig1}(a), the curve corresponding to the 
SSE experiments lies significantly below the reference one. This shows that when 
the system is left unperturbed at constant temperature $T_s$ it rearranges its spin 
configuration toward the equilibrium one for this temperature. The equilibrated state becomes 
frozen in on further cooling, and is retrieved on re-heating. In other terms, the system 
remembers its age or shows a memory phenomenon. It should be mentioned that the weak magnetic 
field only acts as a weak perturbation and has basically no influence on the intrinsic 
equilibration (aging) of the spin configuration at $T_s$. 
Similar effects would be 
obtained for example with an ac-field\cite{acmemo} (as in conventional memory experiments) 
or when keeping a non-zero magnetic field when performing the stop\cite{Ito2}. This will also 
be evidenced in the relaxation measurements described below. Fig.~\ref{fig1}(b) presents the 
difference curves obtained by subtracting from the reference the SSE curves obtained for 
different stop times $t_s$. The stop times are chosen logarithmically spaced, and one 
immediately observes that the minimum of the different curves are closely equidistant from 
each other in magnitude, indicating logarithmic relaxation. The advantage of the dc-method to 
expose fundamental SG features in this simple way has also been demonstrated for other 
glassy systems\cite{nam}.\\
The effects of two stops during cooling are now studied. Fig.~\ref{fig2} shows the results of 
DSE at $T_{s1}$=21K and $T_{s2}$=17K. The main frame shows the difference plots for the DSE, 
as well as the corresponding SSE. The inset depicts the original ZFC magnetization curves 
before subtraction from the reference. Comparing the DSE curve with the corresponding SSE 
ones in the main frame, it is clear that the relaxation at $T_{s1}$=21K is not affected by 
the second stop at lower temperature and reciprocally, the aging at $T_{s2}$ is not influenced 
by the first equilibration at the higher temperature. This illustrates again the memory and 
temperature chaos features. During the first stop, the system reaches a characteristic age - 
or length scale. This age is conserved on further cooling, even if the system is again aged 
at lower temperature toward its equilibrium state at this temperature\cite{addendum}. 
The state equilibrated at the higher temperature thus survives the spin re-configuration 
occurring at lower temperature on shorter length scales, and the system remembers its initial high temperature state on 
re-heating. Of course, if the stop at lower temperature is performed for a much longer time, 
allowing the correlations to develop on longer length scales, a partial re-initialization of 
the spin configuration would occur, unmasking the two different length scales characteristic 
of the two aging processes\cite{memo}. It is of interest to note that recent theoretical 
work\cite{hajime} based on the droplet picture in a chaotic landscape can reproduce the 
experimentally observed rejuvenation and memory phenomena.\\
For comparison, The relaxation of the ZFC magnetization is investigated using the same cooling 
protocols as in the $M$ vs. $T$ measurements. Fig.~\ref{fig3} shows the relaxation curves 
obtained at $T_m$=19K (to compare with 
the single stop depicted in Fig.~\ref{fig1}(a)). The different curves are measured in an 
applied field of 2 Oe according to the following protocols:  \textbf{1}, no stop was made 
during cooling to the lowest temperature and the magnetic field was 
applied 10 s after $T_m$ was reached on re-heating; \textbf{2}, additionally to procedure \textbf{1} a stop 
was made for 3000 s at $T_m$ during cooling and \textbf{3}, a conventional relaxation 
experiment at $T_m$ employing a wait time of 3000 s before applying the field.  One notices 
that at observation times of $\sim$10-30s (corresponding to the effective observation time of 
the magnetization measurement on heating) and longer, curve \textbf{2} lies significantly below 
curve \textbf{1} as is also the case for the SSE magnetization compared to the no stop magnetization in Fig.~\ref{fig1}(a)). It also shows that the SSE curve is almost 
indistinguishable from the curve recorded on a 3000s old system in 
a ``conventional'' ZFC relaxation measurement (curve \textbf{3}). To illustrate the relative 
non-importance of the measurement field, the relaxation 
measurements illustrated in Fig.~\ref{fig3}(a) are repeated, but cooling the system in the field $h_i$=1 Oe, and recording the magnetization with $h_i$+2 Oe=3 Oe. The additional field cooled 
(FC) relaxation is small\cite{fcrelax}, and as seen in Fig.~\ref{fig3}(b), the results are 
virtually unaffected by the presence of an initial field during the cooling and equilibration 
time.\\
The DSE shown in Fig.~\ref{fig2} can also be mimicked by relaxation measurements, as 
illustrated in Fig.~\ref{fig4}. Curve \textbf{1} and \textbf{2} (resp \textbf{1b} and \textbf{2b}) correspond to SSE at (a)$T_s$=$T_m$=21K and (b)$T_s$=$T_m$=17K. Curve \textbf{3} 
and \textbf{3b} correspond instead to the DSE case, with stops at $T_{s1}$=21K and $T_{s2}$=17K, measuring the relaxation at (a)$T_m$=21K or (b)$T_m$=17K. As indicated in the $M$ vs. $T$ 
measurements, the relaxation after the SSE and DSE protocols are virtually identical, 
illustrating again chaos and memory characteristics of the spin glass phase.\\
Note on the relaxation measurements. Looking again at Fig.~\ref{fig2}, one notices that the 
relaxation 'dips' created in the $\Delta M$ vs. $T$ curves are rather broad, which makes 
it difficult to repeat DSE with closer stop temperatures. In that case, the relaxation 
experiments which are here used mainly to confirm the temperature dependent magnetization 
measurements, would yield more information on the interference and overlap 
effects\cite{vincent_djur}.

\section{Conclusion}

Using simple dc-magnetization measurements, performed after specific cooling protocols, memory 
and temperature chaos features are revealed in an Ising spin glass. The results are confirmed 
by relaxation measurements. It would be of interest to overcome the large\cite{Hajimes} 
overlap observed in Monte Carlo simulations and reproduce these results by studying the correlation function at some temperature, after equilibration at a higher one.  

\acknowledgements
This work was financially supported by The Swedish Natural Science Research Council (NFR).

\newpage

\end{multicols}


\begin{references}
\bibitem{addendum}Again, due to the chaotic nature of the SG phase, the equilibrium state at the lower temperature is distinct from the higher temperature one, so that the relaxation at the lower temperature is not affected by the first equilibration at the higher one.
\end{references}
\end{document}